\documentclass{article}

\usepackage{arxiv}

\usepackage[utf8]{inputenc} 
\usepackage[T1]{fontenc}    
\usepackage{hyperref}       
\usepackage{url}            
\usepackage{booktabs}       
\usepackage{amsfonts}       
\usepackage{nicefrac}       
\usepackage{microtype}      
\usepackage{lipsum}		
\usepackage{graphicx}
\usepackage{doi}
\usepackage{bm}
\usepackage{multirow}
\usepackage{siunitx}
\usepackage[english]{babel}
\usepackage{amsmath,amssymb,amstext}
\numberwithin{equation}{section}
\usepackage{color}
\usepackage[backend=biber, style=nature, url=false,isbn=false]{biblatex}
\DeclareSourcemap{
	\maps[datatype=bibtex]{
		\map{
			\step[fieldset=abstract, null]
		}
	}
}

\title{Local statistical moments to capture Kramers-Moyal coefficients}

\author{
	Christian Wiedemann$^{1, 2, *}$, Matthias Wächter$^1$, Jan A. Freund$^2$, Joachim Peinke$^1$ \\
	\and
	$^1$ \\
	School of Mathematics and Science, ForWind \\
	Carl von Ossietzky University\\
	Oldenburg, Germany \\
	\and
	$^2$ \\
	School of Mathematics and Science, ICBM \\
	Carl von Ossietzky University\\
	Oldenburg, Germany \\
	\and
	$*$ \\
	\texttt{christian.wiedemann@uni-oldenburg.de}\\
}

\renewcommand{\shorttitle}{Local statistical moments}

\hypersetup{
	pdftitle={Title},
	pdfsubject={},
	pdfauthor={Christian~Wiedemann},
	pdfkeywords={},
}

\begin{document}
	
	
	\thispagestyle{empty}
	\newpage
	
	\maketitle
	\setcounter{page}{1}

	\begin{abstract}
		This study introduces an innovative local statistical moment approach for estimating Kramers-Moyal coefficients, effectively bridging the gap between nonparametric and parametric methodologies. These coefficients play a crucial role in characterizing stochastic processes. Our proposed approach provides a versatile framework for localized coefficient estimation, combining the flexibility of nonparametric methods with the interpretability of global parametric approaches. We showcase the efficacy of our approach through use cases involving both stationary and non-stationary time series analysis. Additionally, we demonstrate its applicability to real-world complex systems, specifically in the energy conversion process analysis of a wind turbine.
	\end{abstract}

	\section{Introduction}
	
	Estimating the Kramers-Moyal coefficients is a fundamental task in the analysis of stochastic processes \cite{risken1996fokker,tabar2019analysis}. These coefficients provide valuable insights into the underlying dynamics of various systems, ranging from financial markets and climate systems to biological processes \cite{friedrich2011approaching,risken1996fokker}. Accurate estimation of these coefficients is crucial for understanding and predicting the behavior of such systems. For instance, in the domain of wind energy, understanding the stochastic behavior of wind turbine performance is crucial for optimizing their design and operational strategies \cite{lin2023discontinuous,lind2014reconstructing,gottschall2008improve,mucke2015langevin,milan2013turbulent}.

	In the past, two main approaches have been employed for estimating Kramers-Moyal coefficients: nonparametric methods \cite{tabar2019analysis,gorjao2019kramersmoyal,rinn2016langevin} and global statistical moments methods \cite{nikakhtar2023data,kleinhans2007maximum,kleinhans2012estimation,garcia2017nonparametric}. Nonparametric methods, such as the Nadaraya-Watson estimator and the binning approach, offer flexibility and can capture intricate dynamics. However, they often suffer from computational inefficiency and lack interpretability, particularly in complex systems.
	
	On the other hand, global statistical moments approaches typically assume stationarity and model the system's dynamics with a single set of parameters. While these methods provide simplicity and interpretability, they may oversimplify the complex behavior of systems and fail to capture localized or non-stationary dynamics accurately. This limitation restricts their applicability in various real-world scenarios.
	
	To overcome the limitations of existing approaches, we propose a novel local statistical moment approach for estimating Kramers-Moyal coefficients. This approach combines the advantages of both nonparametric and global statistical moments methods, offering a versatile framework for localized estimation. Through our innovative approach, we compute multiple parameter sets, each coupled with a set of functions detailing the local dynamics of the system.

	The main focus of this study is to introduce and validate the effectiveness of our innovative local statistical moment approach for estimating Kramers-Moyal coefficients. To accomplish this, we employ distinct datasets: simulated stationary data, simulated non-stationary data and real-world wind turbine data. In addition to presenting our new approach, we also provide a comprehensive overview of existing nonparametric and global statistical moments methods.
	
	
	
	
	\section{Estimation of the Kramers-Moyal coefficients}
	
	In this section, first, we will provide an overview of two established approaches for capturing the n'th conditional moment: a nonparametric approach and a global statistical moments approach \cite{nikakhtar2023data}. Additionally, we will introduce a novel technique using a local statistical moment approach to estimate the n'th conditional moment. To maintain simplicity and focus, we will illustrate these methods in the context of a one-dimensional time series. However, it is worth noting that each of these approaches can be extended to handle systems of coupled stochastic differential equations.
	
	To provide a foundation for our analysis, we begin by introducing a one-dimensional non-stationary Langevin equation. This equation serves as a basis for understanding the time evolution of the state variable, denoted as $x(t)$, under the fulfillment of a given condition represented by the parameter $\vec{g}$ (grid point). The equation can be expressed as \cite{risken1996fokker,tabar2019analysis}:
	
	\begin{align}
		\dot{x}(t)|_{\vec{c}(t)=\vec{g}} = D^{(1)}(\vec{g}) + \sqrt{D^{(2)}(\vec{g})} \cdot \Gamma(t)
	\end{align}
	
	The term $\Gamma(t)$ represents a zero-mean delta correlated Gaussian white noise, with an autocorrelation given by $\langle \Gamma(t) \cdot \Gamma(t') \rangle = 2 \cdot \delta(t-t')$. This noise follows a Gaussian distribution $\Gamma \sim \mathcal{N}(0, 2)$ with an intensity of 2. It introduces stochasticity into the equation, accounting for random fluctuations on the evolution of $x(t)$.
	
	The parameter $\vec{g}$ in the equation serves as a flexible placeholder for various conditions, such as time or a specific point in state space. For instance, $\vec{g}$ can be represented as a vector $(x_0, y_0, t_0, \dots)$, encapsulating specific states of variables like $x$ and $y$, along with a corresponding timestamp.
	
	
	The n'th conditional moments of a time series $x$, denoted as $M^{(n)}(\vec{g}, \tau)$, are defined as \cite{risken1996fokker,tabar2019analysis}: 
	
	\begin{align}
		M^{(n)}(\vec{g}, \tau) &= \langle \left(x(t+\tau)-x(t) \right)^{n} | \vec{c}(t) = \vec{g} \rangle
	\end{align}
	
	These moments capture the statistical properties of the difference between the values of $x$ at time $t$ and $t+\tau$, conditioned on the condition $\vec{c}(t)$ being equal to $\vec{g}$. We will call these differences increments, with $\Delta_{\tau} x(t) = x(t+\tau)-x(t)$.
	
	The Kramers-Moyal coefficients, denoted as $D^{(n)}(\vec{g})$, can be calculated using these conditional moments \cite{risken1996fokker,tabar2019analysis}:
	
	\begin{align}
		D^{(n)}(\vec{g}) &= \lim_{\tau \rightarrow 0} \frac{M^{(n)}(\vec{g}, \tau)}{n! \cdot \tau} \label{eq:moment_to_kmc}
	\end{align}
	
	The equations provided above represent the analytical approach for defining the n'th conditional moments, denoted as $M^{(n)}(\vec{g}, \tau)$, and the calculation of the corresponding Kramers-Moyal coefficients, denoted as $D^{(n)}(\vec{g})$. 
	
	However, in practical scenarios, it is often necessary to estimate these conditional moments using numerical data rather than having direct analytical expressions. Therefore, alternative approaches are employed to estimate the conditional moments.
	
	Next, we explore the typical approaches employed to capture the n'th conditional moment and introduce our novel local statistical moments approach. 

	
	In our analysis, we consider a discrete time series denoted as $x_i$, where $i$ represents the index of the data point. Additionally, we have a corresponding condition time series denoted as $\vec{c}_i$. Both time series are assumed to have a constant sampling frequency $\frac{1}{\Delta t}$, meaning that $x_i$ corresponds to the value of $x$ at time $i \cdot \Delta t$. The length of each time series is denoted as $N$, representing the total number of elements in the series. This implies that $x_i$ and $\vec{c}_i$ have $N$ data points.
	
	To calculate the increments over a specific time period, we introduce the parameter $\tau_m$, defined as $\tau_m = m \cdot \Delta t$. Here, $m$ represents the number of time steps and $\Delta t$ is the time interval between consecutive data points. The increment at time $i$ over a period of $\tau_m$ is denoted as $\Delta_{\tau_m} x_i$ and can be calculated as $\Delta_{\tau_m} x_i = x_{i+m} - x_i$.
	
	\subsection{Nonparametric estimator}
	
	The two most commonly used nonparametric estimators for the conditional moments are the Nadaraya-Watson estimator \cite{nadaraya1964estimating,watson1964smooth,lamouroux2009kernel} and the binning approach \cite{rinn2016langevin}.
	
	The Nadaraya-Watson estimator is a versatile method that utilizes kernel functions and a given bandwidth to calculate weights for each condition. These weights determine the influence of neighboring data points on the estimation of the conditional moments. By assigning higher weights to nearby data points and lower weights to more distant ones, the Nadaraya-Watson estimator captures the local dynamics.
	
	The binning approach is a specific case of the Nadaraya-Watson estimator. It involves restricting the estimation to grid points and their corresponding bandwidths. By dividing the range of conditions into discrete, non-overlapping bins, the binning approach calculates the conditional moments by averaging the increments within each bin with equally distributed weights. 
	
	To implement the Nadaraya-Watson estimator, we first introduce a procedure for calculating the weights $\omega_{\vec{h}}(\vec{x})$ using kernel functions $k_i$
	
	\begin{align}
		\vec{x} = \left( \begin{array}{c}
			x_1 \\
			x_2 \\
			\vdots \\
			x_D 
		\end{array} \right); \qquad 
		\vec{h} = \left( \begin{array}{c}
			h_1 \\
			h_2 \\
			\vdots \\
			h_D 
		\end{array} \right); \qquad
		K_{\vec{h}}(\vec{x}) = \prod_{d=1}^{D} k_i\left(\frac{x_d}{h_d}\right) \label{eq:kernel}
	\end{align}
	
	and a given bandwidth $\vec{h}$ specific to each condition. These weights
	
	\begin{align}
		\omega_{\vec{h}}(\vec{x}_i) &= \frac{K_{\vec{h}}(\vec{x}_i)}{\sum_{i=1}^N K_{\vec{h}}(\vec{x}_i)} \label{eq:weights}
	\end{align}
	
	determine the contribution of individual data points to the estimation of the conditional moments. Then, we apply the Nadaraya-Watson estimator to obtain the estimated conditional moments, incorporating the calculated weights and the available data. In Table \ref{tab:kernel}, we provide a demonstration of some of the most commonly used kernel functions in the literature. These kernel functions include the Gaussian kernel, the Epanechnikov kernel, and the rectaangular kernel. Each kernel function has a specific shape and bandwidth parameter that can be adjusted to fit the characteristics of the data and the desired estimation goals.


	\newcolumntype{L}[1]{>{\raggedright\let\newline\\\arraybackslash\hspace{0pt}}m{#1}}

	\begin{table}
		\caption{This list showcases a selection of not normalized kernel functions commonly used in various estimation methods with bandwidth $h$. Each kernel function exhibits different properties and characteristics, making them suitable for different analysis scenarios.}
		\centering
		\begin{tabular}{L{0.07\linewidth}L{0.25\linewidth}L{0.25\linewidth}L{0.28\linewidth}L{0.25\linewidth}}
			\toprule
			Kernel & Efficiency & $k_h(x)$\\
			\midrule
			\addlinespace[15.5pt]
			\raisebox{-.1\normalbaselineskip}[0pt][0pt]{\rotatebox[origin=c]{90}{Epanechnikov}}  & considered efficient, providing accurate estimates with the same amount of data & $\begin{cases}
				1-\left(\frac{x}{h}\right)^2 , & \text{if } |x| \leq 1.0/h\\
				0 , & \text{if } |x| > 1.0/h
			\end{cases} $ \\
			\addlinespace[15.5pt]
			\midrule
			\addlinespace[12.5pt]
			\raisebox{-.1\normalbaselineskip}[0pt][0pt]{\rotatebox[origin=c]{90}{Gaussian}}  &  mathematically tractable and computationally efficient, especially in high-dimensional problems & $\exp(-0.5 \cdot \left( \frac{x}{h} \right)^2)$ \\ \addlinespace[5pt]
			\midrule
			\addlinespace[15pt]
			\raisebox{-.1\normalbaselineskip}[0pt][0pt]{\rotatebox[origin=c]{90}{Rectangular}} & computationally efficient due to its simple shape and properties &  $\begin{cases}
				1 , & \text{if } |x| \leq h\\
				0 , & \text{if } |x| >h
			\end{cases}$ \\
			\addlinespace[15pt]
			\bottomrule
		\end{tabular}
		\label{tab:kernel}
	\end{table}
	
	The conditional moments can be estimated with these weights $\omega_{\vec{h}}(\vec{x})$:
	
	\begin{align}
		\hat{M}^{(n)}(\vec{g}, \tau) &= \sum_{i=1}^{N} \left( \Delta_{\tau_m} x_i \right)^n \cdot \omega_{\vec{h}}(\vec{c}_i-\vec{g}) \label{eq:nadaraya}
	\end{align}
	
	In this equation, $\hat{M}^{(n)}(\vec{g}, \tau_m)$ represents the estimated n'th conditional moment for a specific condition $\vec{g}$ and time lag $\tau_m$. The estimator is calculated by averaging the n'th powers of the increments over a the time lag $\tau_m$, weighted by $\omega_{\vec{h}}(\vec{c}_i-\vec{g})$. The multidimensional kernel function $K_{\vec{h}}(\vec{c}_i-\vec{g})$ assigns weights to each data point based on the difference between the condition $\vec{c}(t)$ and the specific condition $\vec{g}$.
	
	We can formulate the Kramers-Moyal coefficients $\hat{D}^{(n)}(\vec{g})$ with \eqref{eq:moment_to_kmc} as it follows.
	
	\begin{align}
		\hat{D}^{(n)}(\vec{g}) =   \lim_{\tau_m \rightarrow 0} \frac{\hat{M}^{(n)}(\vec{g}, \tau_m)}{\tau_m \cdot n!}
	\end{align}

	In the binning approach, the grid points are defined to span the range of values in each condition. 
	
	The bandwidth in the binning approach is determined by the distance between two adjacent grid points. For the $j$'th bandwidth, denoted as $h_j$, it is calculated as the difference between the values of two consecutive elements of $\vec{g}_j$. Mathematically, $h_j = g_{j, k+1} - g_{j, k}, \forall k <  \tilde{N}_j$.
	
	In the binning approach, the rectangular kernel in combination with the setup described earlier is used with \eqref{eq:nadaraya}. The rectangular kernel assigns equal weight to all data points within a specific bin. Every data point is exclusively assigned to one bin based on the condition values. This means that each data point contributes to the calculation of the conditional moments for a single bin and does not influence the estimation of other bins.
	
	\subsection{Global statistical moments approach}
	
	The global statistical moments approach involves estimating a set of coefficients denoted as $\vec{\phi}^{(n)}(\tau_m)$ with a given set of functions represented by $\vec{f}^{(n)}(x)$. These coefficients and functions are used to model the n'th conditional moment
	
	\begin{align}
		\tilde{M}^{(n)}(x, \tau_m) &= \sum_{j=1}^{N_f} \phi_j^{(n)}(\tau_m) \cdot f_j^{(n)}(x).
	\end{align}
	
	For a comprehensive understanding of the details and intricacies of the global statistical moments approach, we refer to the publication \cite{nikakhtar2023data}. This source provides a comprehensive description of the methodology.
	
	We can formulate the Kramers-Moyal coefficients $\tilde{D}^{(n)}(x, \vec{g})$ as it follows.
	
	\begin{align}
		\tilde{D}^{(n)}(x, \vec{g}) = \sum_{j=1}^{N_f} \Phi_j^{(n)}(\vec{g}) \cdot f_j(x),
	\end{align}
	
	With \eqref{eq:moment_to_kmc} we can calculate the coefficients with $\phi_j^{(n)}(\vec{g}, \tau_m)$.
	
	\begin{align}
		\Phi_j^{(n)}(\vec{g}) = \lim_{\tau_m \rightarrow 0} \frac{\phi_j^{(n)}(\vec{g}, \tau_m)}{\tau_m \cdot n!}
	\end{align}
	
	Although the functions $\vec{f}$ can depend on multiple variables in practice, for the purpose of presenting the method here, we limit our discussion to the one-dimensional representation. In this context, we will refer to these functions as fit-functions. However, it is important to note that these functions can, in principle, be arbitrary and flexible, allowing for the incorporation of various mathematical forms and functional relationships. An example of a set of functions can be a polynomial of order $K$. The vector of functions $\vec{f}(x)$ can be defined as:
	
	\begin{align}
		\vec{f}(x) = (1, x, x^2, \dots, x^K)
	\end{align}
	
	In this case, the functions in the vector $\vec{f}(x)$ are powers of $x$ ranging from $x^0$ to $x^K$. 
	
	To facilitate the estimation process within the global statistical moments approach, we introduce the function matrix denoted as 
	\begin{align}
		\bm{F}^{(n)}_{i, j}(x) = f_i(x) \cdot f_j(x). \label{eq:matrix_f}
	\end{align}
	
	This matrix encapsulates the evaluation of the functions $f_i$ for the data point $x$.
	
	With this setup, the coefficients $\vec{\phi}^{(n)}(\tau_m)$ can be estimated using the following equation:
	
	\begin{align}
		\vec{\phi}^{(n)}(\tau_m) = \left( \frac{1}{N} \sum_{i=1}^N \bm{F}^{(n)}(x_i) \right)^{(-1)} \cdot \left( \frac{1}{N} \sum_{i=1}^N \Delta_{\tau_m} x_i \cdot \vec{f}^{(n)}(x_i) \right)
	\end{align}
	
	In this equation, $\bm{F}^{(n)}(x_i)$ represents the function matrix evaluated at $x_i$. The term $\Delta_{\tau_m} x_i$ denotes the increment in the time series $x_i$ over the time period $\tau_m$, and $\vec{f}^{(n)}(x_i)$ corresponds to the vector of functions evaluated at $x_i$.
	
	The estimation of the coefficients involves averaging the product of the increments $\Delta_{\tau_m} x_i \cdot \vec{f}^{(n)}(x_i)$ over all data points and multiplying it by the inverse of the average function matrix $\bm{F}^{(n)}(x_i)$.
	
	\subsection{Local statistical moment approach}
	
	The local statistical moment approach involves estimating a set of coefficients, denoted as $\vec{\phi}^{(n)}(\vec{c}, \tau_m)$, for each grid point $\vec{g}$ using a given set of functions represented by $\vec{f}^{(n)}(\vec{d})$, where $\vec{d}$ is the dependency variable. This allows us to model the n'th conditional moment as follows:
	
	\begin{align}
		\check{M}^{(n)}(\vec{d}, \vec{c}, \tau_m) &= \sum_{j=1}^{N_f} \phi_j^{(n)}(\vec{c}, \tau_m) \cdot f_j(\vec{d})
	\end{align}
	
	In this context, it is important to differentiate between conditions $\vec{c}$ and dependencies $\vec{d}$, where both are subsets of the grid point $\vec{g}$. The coefficients $\vec{\phi}^{(n)}(\vec{c}, \tau_m)$ depend on the conditions $\vec{c}$, while the functions generally $\vec{f}^{(n)}(\vec{d})$ depend on the dependencies.
	
	For the sake of simplicity, we limit our discussion to a one-dimensional dependency represented by $x$. Additionally, we refer to the conditions $\vec{c}$ as the grid point $\vec{g}$. This simplification allows for a clearer understanding of the approach while maintaining comprehensibility. 
	
	\begin{align}
		\check{M}^{(n)}(x, \vec{g}, \tau_m) &= \sum_{j=1}^{N_f} \phi_j^{(n)}(\vec{g}, \tau_m) \cdot f_j(x)
	\end{align}
	
	To estimate the coefficients, we utilize the weights defined in Equations \eqref{eq:kernel} and \eqref{eq:weights}, and the function matrix as defined in Equation \eqref{eq:matrix_f}. These components enable us to estimate the coefficients $\vec{\phi}^{(n)}(\vec{c}, \tau_m)$ using the following equation:
	
	\begin{align}
		\vec{\phi}^{(n)}(\vec{g}, \tau_m) &=  \left( \sum_{i=1}^N \omega_{\vec{h}}(\vec{c}_i-\vec{g}) \cdot \bm{F}^{(n)}(x_i) \right)^{(-1)} \cdot \left(\sum_{i=1}^N \omega_{\vec{h}}(\vec{c}_i-\vec{g}) \cdot\Delta_{\tau_m} x_i \cdot \vec{f}^{(n)}(x_i) \right) \label{eq:local_parametric}
	\end{align}
	
	We can formulate the Kramers-Moyal coefficients $\check{D}^{(n)}(x, \vec{g})$
	
	\begin{align}
		\check{D}^{(n)}(x, \vec{g}) = \sum_{j=1}^{N_f} \Phi_j^{(n)}(\vec{g}) \cdot f_j(x).
	\end{align}
	
	With \eqref{eq:moment_to_kmc} we can calculate the coefficients with $\phi_j^{(n)}(\vec{g}, \tau_m)$ 
	
	\begin{align}
		\Phi_j^{(n)}(\vec{g}) = \lim_{\tau_m \rightarrow 0} \frac{\phi_j^{(n)}(\vec{g}, \tau_m)}{\tau_m \cdot n!}.
	\end{align}
	
	In contrast to the global statistical moments approach, the local statistical moment approach employs a weighted average instead of a simple average. The weights used in the weighted average are dependent on the grid point $\vec{g}$, the bandwidths $\vec{h}$, and the kernel functions $k$.
	
	These weights take into account the specific characteristics of each grid point, as determined by the conditions $\vec{c}$, and are influenced by the corresponding bandwidths $\vec{h}$ and kernel functions $k$. By incorporating these weights, the local statistical moment approach assigns higher importance to data points that are closer to the grid point, while giving less weight to points farther away.
	
	Indeed, we can argue that the global statistical moments approach can be seen as a subset of the local statistical moment approach, where all data points are assigned equal weights. In this case, the weights are constant, given by $\omega_{\vec{h}}(\vec{c}_i-\vec{g}) = \frac{1}{N}$, resulting in a simple average.
	
	Similarly, the nonparametric approach can be considered as a subset of the local statistical moment approach, where the set of functions $\vec{f}(x)$ is reduced to a constant function. This constant function effectively ignores the dependency variable, resulting in a nonparametric estimation solely based on the conditional moments.
	
	Therefore, the local statistical moment approach encompasses both the global statistical moments approach and the nonparametric approach, making it a more general and versatile framework. It unites these two approaches by allowing for flexible choices of weights and functions, enabling localized estimation of the conditional moments.
	
	By incorporating both weighted averaging and a diverse set of functions, the local statistical moment approach leverages the advantages of both the global statistical moments and nonparametric approaches. It can capture localized dynamics while maintaining interpretability and flexibility in the estimation process.
	
	From a numerical perspective, the local statistical moment approach offers significant advantages. It provides a unified framework that allows for the estimation of local statistical moment coefficients while also accommodating the nonparametric and global statistical moments approaches.

	\section{Numerical simulations} \label{sec:numerical}
	
	In this section, we present numerical simulations to demonstrate the effectiveness of the estimation methods for Kramers-Moyal coefficients in different scenarios. We focus on two aspects: the analysis of stationary one-dimensional processes using all three mentioned approaches and the examination of synthetic non-stationary processes using our new local statistical moment approach.
	
	To showcase the key differences in the estimation methods, we consider two stationary one-dimensional processes. By comparing the results obtained from the nonparametric, global statistical moments, and local statistical moment approaches, we can gain insights into their strengths and limitations in capturing the Kramers-Moyal coefficients accurately.
	
	In addition to stationary processes, we investigate synthetic non-stationary processes to highlight the capabilities of our new local statistical moment approach. These non-stationary processes exhibit time-varying dynamics and present challenges for traditional estimation methods. By employing our novel approach, we aim to demonstrate its ability to capture localized dynamics and provide improved modeling accuracy for non-stationary systems. Through these synthetic examples, we illustrate how the local statistical moment approach enhances our understanding and analysis of non-stationary processes. 
	
	All simulations are done with the Euler-Maruyama scheme
	
	\begin{align}
		x_{i+1} = x_i + dt \cdot D^{(1)}(x_i) + \sqrt{dt \cdot D^{(2)}(x_i)} \cdot \Gamma_i.
	\end{align}
	
	Here, $x_0 = 0$ is the starting point, $N = 10^5$ is the number of elements and $dt = 0.1$ is the time step to generate the time series.
	
	We recognize that each method has the potential for improved performance. Our focus was not on fine-tuning the estimation methods for optimal results; rather, we aimed to provide a concise overview and comparison of the methods.
	
	\subsection{Stationary one-dimensional processes}
	
	In this section, we introduce two distinct one-dimensional processes and conduct a comparative analysis of the estimation results using various methods.
		
	For both our test cases here, we use the following weights, kernels and fit-functions. For the estimation of the weights, we use for the nonparametric and local statistical moment approach a gaussian shaped kernel with a bandwidth of $0.25$. As a fit function we use a linear function for the global statistical moment approach and the local statistical moment approach.
	
	\begin{align}
		\omega(x-x_0) &= \exp\left(-\frac{1}{2} \cdot \left(\frac{x-x_0}{0.5}\right)^2 \right) \text{: Gaussian kernel with bandwidth of 0.5}\\
		\vec{f}(x) &= (1, x) \text{: Linear function with dependency on x}
	\end{align}
	.
	
\begin{figure}
		\centering
		\includegraphics[width=0.95\textwidth]{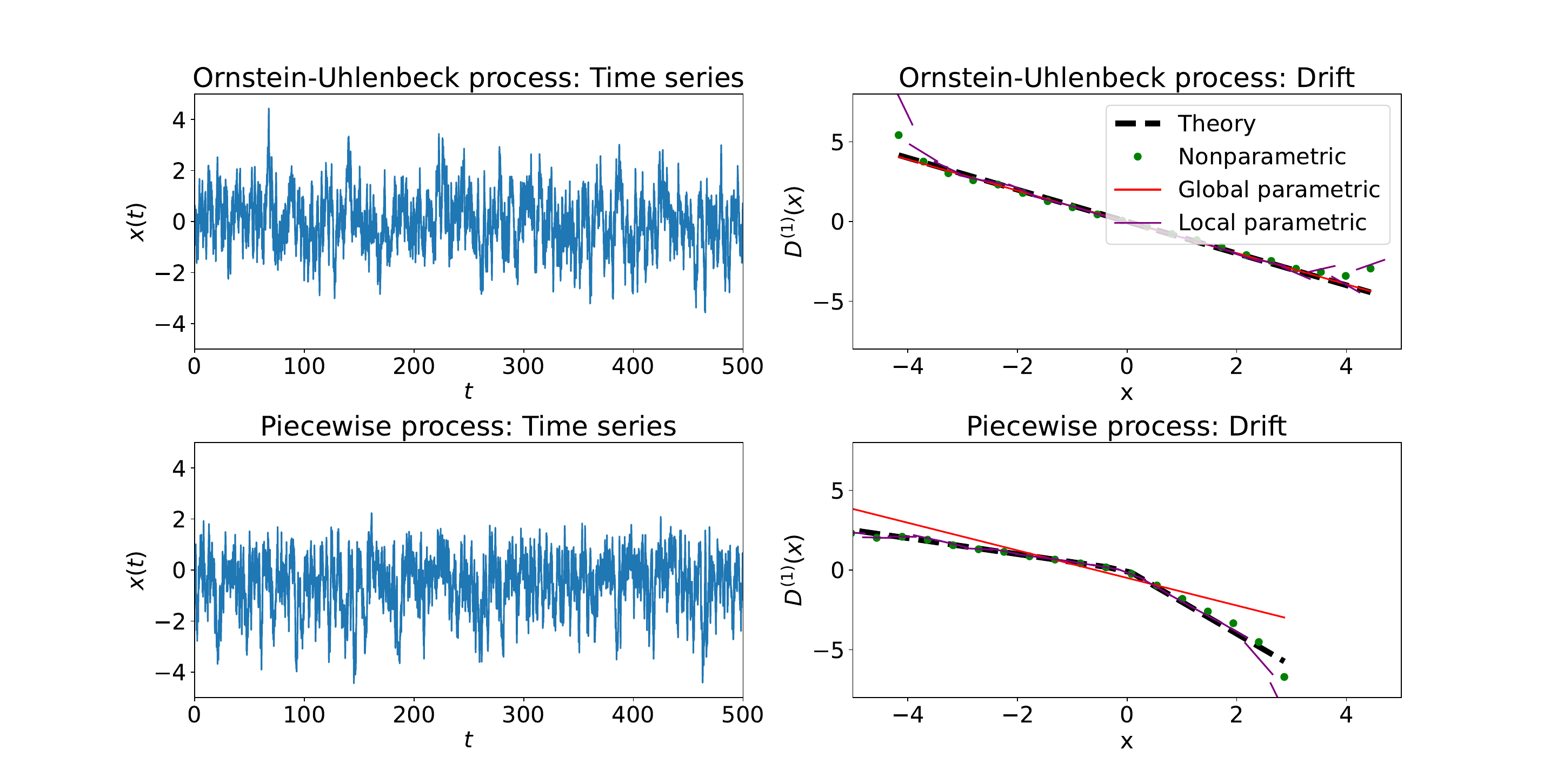}
		\caption{Time series plots on the left depict the Ornstein-Uhlenbeck process (top) and the Piecewise process (bottom). On the right, corresponding drifts are illustrated, including the theoretical drift alongside the estimated drifts obtained through three distinct methods: nonparametric, global statistical moments, and local statistical moments.}
		\label{fig:ou_1}
	\end{figure}

	The first simulation we conducted involved an Ornstein-Uhlenbeck process,

	\begin{align}
		\dot{x}(t) &= -x(t) + \Gamma(t) \label{eq:ou_process_1}
	\end{align}
	
	a classic model used in various disciplines to study the behavior of systems subjected to stochastic fluctuations and mean-reverting tendencies. Upon examining the outcomes of this simulation, depicted in figure \ref{fig:ou_1} (top), a noteworthy revelation surfaced: the global estimation method consistently surpassed alternative approaches in terms of mean absolute distance to the actual drift value. Particularly at the extremities ($|x|>3.5$), both the nonparametric approach and the local statistical moment approach exhibited growing inaccuracies, which correlated with the lower density of the time series towards the outer regions. This discovery underscores the efficacy of the global estimation method, especially when an accurate, known function for the dynamics (in this instance, a linear function) is available.

	The second simulation featured a piecewise linear function (we will refer to this as the Piecewise process), 
	
	\begin{align}
			 \dot{x}(t) &= \Gamma(t) + \begin{cases} 
			-0.5 \cdot x(t) & ,x(t) \leq 0 \\
			-2 \cdot x(t) & ,x(t) > 0 
		\end{cases} \label{eq:ou_process_2}
	\end{align}
	
	a model that differs from the Ornstein-Uhlenbeck process. An illustrative time series example and the outcomes regarding the estimation of the drift coefficient are presented in Figure \ref{fig:ou_1} (bottom). In this instance, the nonparametric and local statistical moment approaches exhibited superior performance. However, it's noteworthy that while the global estimation method didn't fare as well under these conditions, we recognize that employing higher-degree polynomials or a piecewise function could have potentially improved its results.
	
	Moreover, it is essential to emphasize that the efficacy of adopting the global estimation method hinges on our understanding of the dynamics specific to the system. We acknowledge that successful adaptations require a comprehensive grasp of the system's behavior, underscoring the significance of incorporating domain knowledge for optimal outcomes.
	
	In this case, the nonparametric and local statistical moment approaches demonstrated superior results without necessitating adjustments. Nevertheless, it's important to acknowledge that the effectiveness of these local methods may vary, and there are situations where adjustments become imperative through the integration of domain knowledge. Additionally, we observe a recurring trend of diminished accuracy towards the edges of the local methods ($x > 2$).
	
	We recognize the inherent variability in the dynamics of different systems, and in certain cases, fine-tuning parameters or integrating domain-specific information may be imperative to optimize the performance of local methods. Therefore, while our current findings highlight the adaptability of the nonparametric and local statistical moment approaches in this specific scenario, we acknowledge the potential necessity for modifications in other systems based on their unique dynamics.

	In summary, our study highlights the importance of selecting the most suitable estimation method based on the characteristics of the system under investigation. It's not a one-size-fits-all scenario, and the choice of method can significantly impact the accuracy of the results. By acknowledging the strengths and limitations of each approach, researchers can better tailor their methods to the specific dynamics of their systems and enhance the quality of their estimations.

	\subsection{Stationary two-dimensional processes}
	
	In another test system, we introduce a coupled two-dimensional stationary process
	
	\begin{align}
		\left( \begin{array}{c}
			\dot{x}(t) \\
			\dot{y}(t)
		\end{array} \right) &= 
		\left( \begin{array}{c}
			- |y(t)| \cdot x(t) + \Gamma_x(t) \\
			- 0.25 \cdot y(t) + \Gamma_y(t)
		\end{array} \right) \label{eq:twodimensional_stationary}
	\end{align}
	
	to further explore the capabilities of our novel estimation approach. The specific objective of this experiment is to estimate the drift of the $x$-component, which is characterized by the function $D_x^{(1)}(x, y) = - |y(t)| \cdot x(t$).
	
	The $y$-component serves as the condition for our estimation, while our estimation targets the $x$-component. The approach involves using a linear function as part of the fitting procedure to capture the relationship between these two components. In this experiment, we employ the following settings for the estimation process: 
	
	\begin{align}
		\omega(y-y_0) &= \exp\left(-\frac{1}{2} \cdot \left(\frac{y-y_0}{0.5}\right)^2 \right) \text{: Gaussian kernel with bandwidth of 0.5}\\
		\vec{f}(x) &= (1, x) \text{: Linear function with dependency on x}
	\end{align}
	
\begin{figure}
		\centering
		\includegraphics[width=0.95\textwidth]{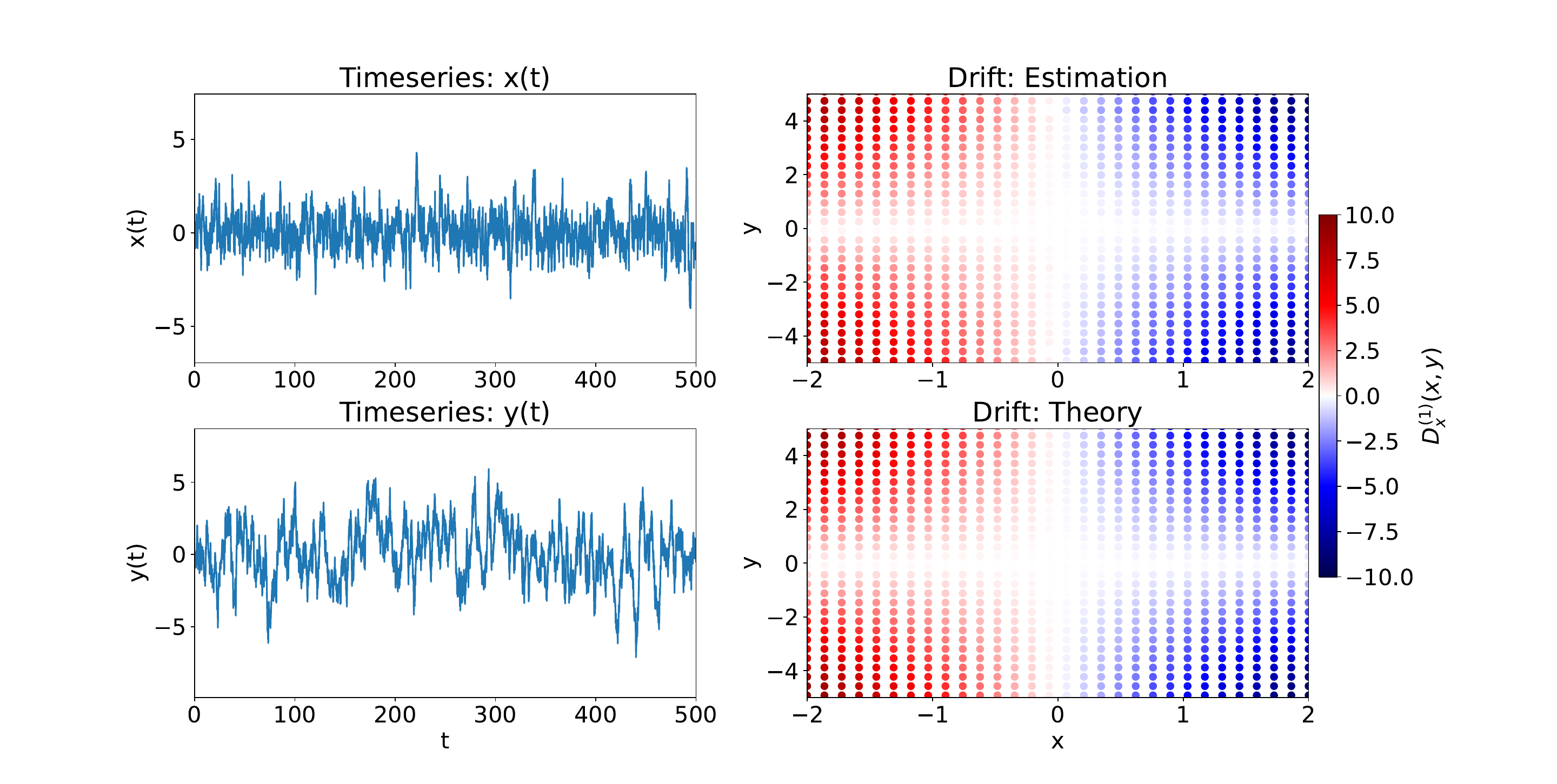}
		\caption{Simulated time series of the $x$ time series (top left), simulated time series of the $y$ time series (bottom, left), the estimation of the drift coefficients (top right) as a scatter plot with color coded drift value and the theoretical drift value (bottom right) of the two dimensional stationary coupled system \eqref{eq:twodimensional_stationary}.}
		\label{fig:twodimensional_stationary}
	\end{figure}

	The results from this test system are shown in figure \ref{fig:twodimensional_stationary}. Our novel approach demonstrates its efficacy by successfully reconstructing the drift function. The ability to condition the estimation on the $y$-component while applying a parameter fit based on the $x$-component allows us to capture the underlying drift dynamics.
	
	\subsection{Non-Stationary two-dimensional processes}
	
	The final numerical example in our study explores a non-stationary two-dimensional coupled system. The system's dynamics can be described by a piecewise-defined coupled system:
	
	\begin{align}
		\left( \begin{array}{c}
			\dot{x}(t) \\
			\dot{y}(t)
		\end{array} \right) &=  \begin{cases} 
			\left( \begin{array}{c}
				- |y(t)| \cdot x(t) + \Gamma_x(t) \\
				- 0.25 \cdot y(t) + \Gamma_y(t)
			\end{array} \right) & , t \leq 5000 \\
			\left( \begin{array}{c}
				- |y(t)| \cdot (x(t)-2) + \Gamma_x(t) \\
				- 0.25 \cdot y(t) + \Gamma_y(t)
			\end{array} \right) & , t > 5000 
		\end{cases} \label{eq:non-stationary}
	\end{align}

	It's worth noting that while the slope of the system remains stationary, the system's fixed points undergo a significant change at a specific time point $t = 5000$.
	
	In our experimental setup, we condition the estimation on both the $y$ and time ($t$) components and utilize a linear function that depends on the $x$-component to capture the evolving dynamics of the system. To facilitate this, we employ the following parameter settings for the estimation process: 
	
	\begin{align}
		\omega(\left( \begin{array}{c}
			y-y_0 \\
			t-t_0
		\end{array} \right)) &= \exp\left(-\frac{1}{2} \cdot \left(\frac{y-y_0}{0.5}\right)^2 \right) \cdot \exp\left(-\frac{1}{2} \cdot \left(\frac{t-t_0}{200}\right)^2 \right) \\ &\text{: Gaussian shaped kernels with bandwidth $h_y=0.5$ and $h_t=200$} \nonumber \\
		\vec{f}(x) &= (1, x) \text{: Linear function with dependency on x}
	\end{align}
	
\begin{figure}
		\centering
		\includegraphics[width=0.95\textwidth]{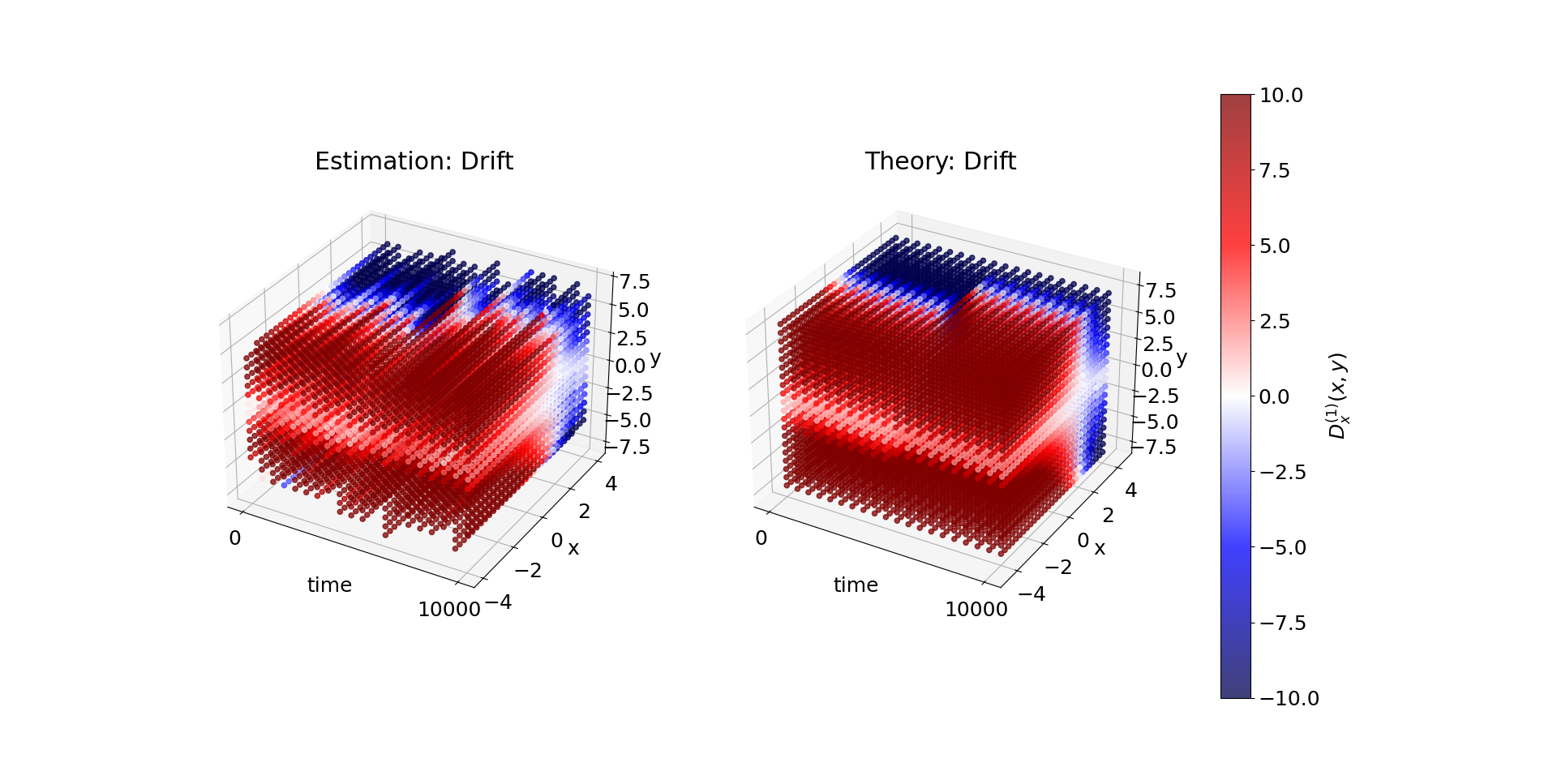}
		\caption{Estimation (left) and theoretical (right) of the drift of the system \eqref{eq:non-stationary}. The drift values are color coded.}
		\label{fig:non-stationary_3d}
	\end{figure}

\begin{figure}
		\centering
		\includegraphics[width=0.95\textwidth]{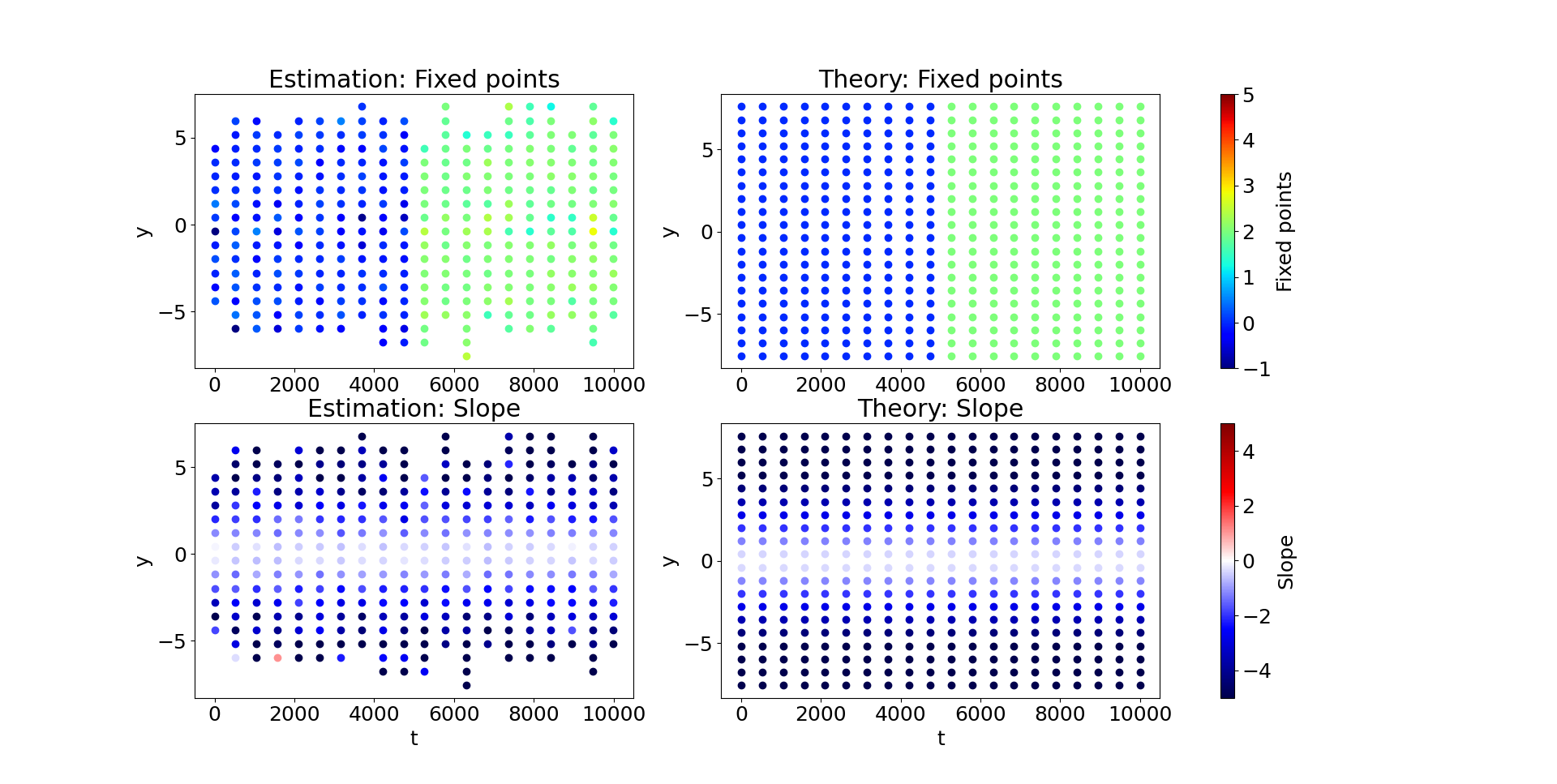}
		\caption{Estimation (left) and theoretical (right) values of the fixed points (top) and the slopes (bottom) of the system \eqref{eq:non-stationary} as color coded values.}
		\label{fig:non-stationary_fixedpoints}
	\end{figure}

	The results obtained from this setup reveal a high degree of accuracy in estimating the dynamics of the system which can be seen in figure \ref{fig:non-stationary_3d} and \ref{fig:non-stationary_fixedpoints}. In figure \ref{fig:non-stationary_3d} we demonstrate the reconstruction of the whole drift map with our new approach, whereas in figure \ref{fig:non-stationary_fixedpoints}, we showcase the estimations of the fixed points and the slope of the drift. Our approach successfully captures the dynamic changes that occur at $t = 5000$, emphasizing the adaptability of our methodology even in the face of non-stationary systems. These findings underscore the effectiveness of our approach in handling time-varying systems.
	
	\section{Real world data}
	
	In a real-world test scenario, we have employed highly-resolved wind turbine SCADA (Supervisory Control and Data Acquisition) data, which offers a detailed account of the turbine's operational parameters over time. Our objective is to employ our novel approach to estimate the non-stationary Langevin power curve over an extensive time period, covering one full year.
	
	The data utilized in this study is sourced from the SCADA system of a Vestas V90 turbine located in the Thanet offshore wind farm. These measurements were recorded at approximately 5-second intervals throughout the year 2017. To ensure consistent time stamps and a stable frequency, the data was aggregated by averaging over 10-second intervals. It is important to note that if no measurements were obtained within the original 5-second interval, the aggregated dataset may contain missing data during the corresponding 10-second interval.

	Our model for the power conversion process is as follows:
	
	\begin{align}
		\dot{P}(t)|_{u(t)=u_0, t = t_0} &= D^{(1)}(P(t), u_0, t_0) + \sqrt{D^{(2)}(P(t), u_0, t_0)} \cdot \Gamma(t)
	\end{align}
	
	Here, $P$ is the power production (in percentage to the rated power) of the turbine, $u$ is the wind speed and $t$ is the time. We model our system with a linear drift along the $P$-axis
	
	\begin{align}
		D^{(1)}(P, u_0, t_0) &= \Phi_0^{(1)}(u_0, t_0) + P \cdot \Phi_1^{(1)}(u_0, t_0)
	\end{align}
	, where the coefficients depend on the wind speed $u$ and the time $t$.
	
	In this study we will focus on the stable fixed points 
	
	\begin{align}
		P_0(u_0, t_0) &= - \frac{\Phi_0^{(1)}(u_0, t_0)}{\Phi_1^{(1)}(u_0, t_0)}\text{ , if } \Phi_1^{(1)}(u_0, t_0)<0.
	\end{align}
	
	For this specific scenario we found a combination of a rectangular kernel function for the $t$-condition and a Epanechnikov kernel for the $u$-condition with the respective bandwidths of $h_u = 0.5$ \si{\meter\per\second} and $h_t = 7$ days to be create meaningful results
	
	\begin{align}
		\omega(\left( \begin{array}{c}
			u-u_0 \\
			t-t_0
		\end{array} \right)) &= \left( \begin{cases}
		1-(\frac{u-u_0}{0.5})^2 , & \text{if } |u-u_0| \leq 0.5\\
		0 , & \text{if } |u-u_0| > 0.5
		\end{cases} \right) \cdot \left(\begin{cases}
	    1 , & \text{if } |t-t_0| \leq 7 \text{days}\\
		0 , & \text{if } |t-t_0| > 7 \text{days}
		\end{cases} \right).
	\end{align}

\begin{figure}
		\centering
		\includegraphics[width=0.95\textwidth]{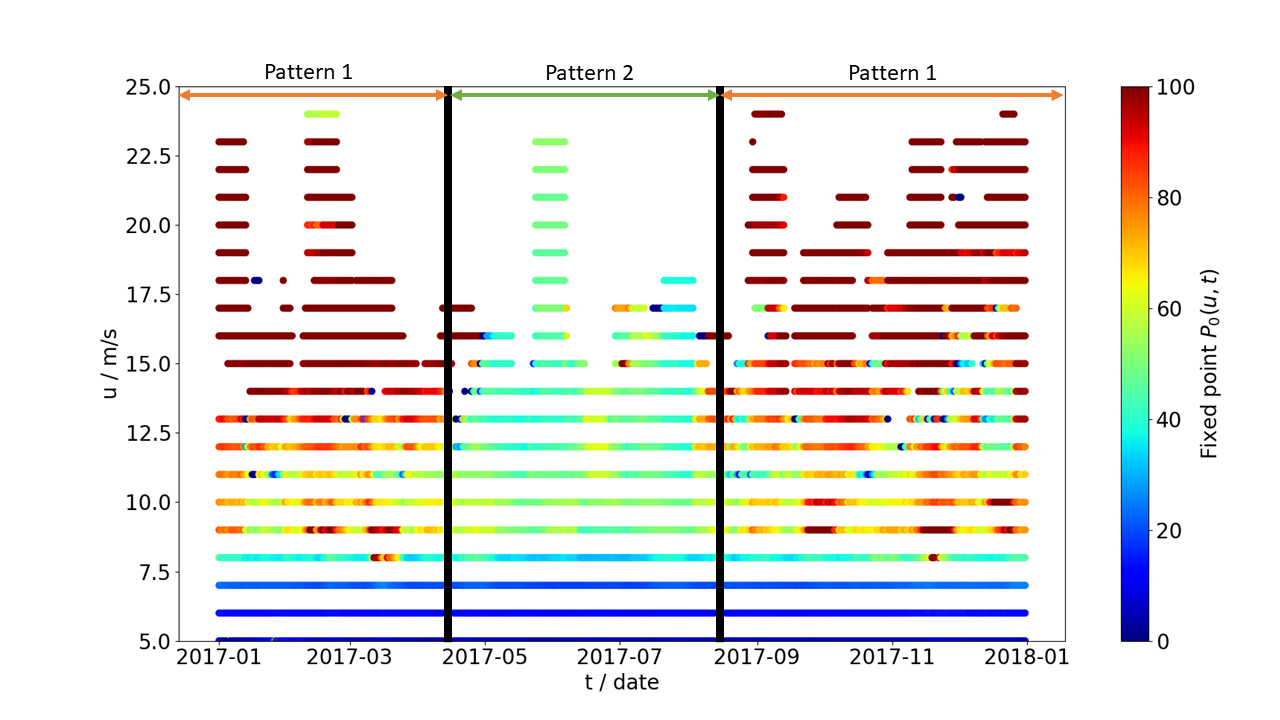}
		\caption{Estimation of the fixed points in terms of percentage of the rated power (color coded) per wind speed (y-axis) of the power conversion process of a wind turbine over a time period of 1 year (x-axis). The black thick lines indicate a change in terms of patterns, which are labeled on top of the plot.}
		\label{fig:powercurve_1year}
	\end{figure}

	We visually represent the results of our analysis in Figure \ref{fig:powercurve_1year}. This figure provides a comprehensive insight into the behavior of the wind turbine power conversion process conditioned on wind speed $u$ and time $t$.
	
	On the x-axis of the graph, we have the dates, creating a time series from the beginning of 2017 to the start of 2018. The y-axis represents wind speed. The key feature of this visualization is the color-coding, which corresponds to the fixed points of power in terms of the percentage of the rated power. The figure reveals two distinct patterns: labeled as pattern 1 $P_1$ and pattern 2 $P_2$. The $P_1$ pattern is dominant from the beginning of 2017 to May 2017, and then again from September 2017 to January 2018. Meanwhile, the $P_2$ pattern emerges during the period between May 2017 and September 2017. The alternating presence of $P_1$ and $P_2$ patterns, which represent variations in the wind turbine's performance, underscores the non-stationary behavior of the system. This dynamic behavior is indicative of the turbine's response to changing conditions, rather than a consistent, steady-state operation. We note that the transition from the $P_1$ pattern to the $P_2$ pattern corresponds with the time period from May 2017 to September 2017. This aligns with the knowledge that the wind turbine was regulated down during this time, a fact substantiated by the observed changes in the power patterns in our results.

	\section{Summary}
	
	In this study, we introduced a novel approach aimed at capturing the dynamics of stochastic systems, utilizing a local statistical moment Ansatz. This approach, positioned between the global statistical moments methods and nonparametric techniques, offers a versatile solution for modeling complex and non-stationary stochastic systems.
	
	To support our novel approach, we provided a comprehensive mathematical foundation, explaining its underlying principles and how it differs from traditional modeling methods. We emphasized the balance it strikes between parameterization and flexibility.
	
	Throughout the study, we presented various numerical examples, highlighting the adaptability and flexibility of our approach. These examples demonstrated its effectiveness in diverse scenarios, showcasing its ability to accurately estimate system dynamics.
	
	In a significant real-world application, we applied our approach to analyze the power conversion process of wind turbines using highly-resolved SCADA data. This case study exemplified our method's practical utility by successfully capturing the dynamics of a real-world system. The results clearly illustrated non-stationary behavior in the wind turbine's power conversion process, underscoring the valuable insights our approach can provide in real-world applications.

	\appendix

	\newpage
	\printbibliography

@article{nikakhtar2023data,
	title={Data-driven reconstruction of stochastic dynamical equations based on statistical moments},
	author={Nikakhtar, Farnik and Parkavousi, Laya and Sahimi, Muhammad and Tabar, M Reza Rahimi and Feudel, Ulrike and Lehnertz, Klaus},
	journal={New Journal of Physics},
	volume={25},
	number={8},
	pages={083025},
	year={2023},
	publisher={IOP Publishing}
}

@article{friedrich2011approaching,
	title={Approaching complexity by stochastic methods: From biological systems to turbulence},
	author={Friedrich, Rudolf and Peinke, Joachim and Sahimi, Muhammad and Tabar, M Reza Rahimi},
	journal={Physics Reports},
	volume={506},
	number={5},
	pages={87--162},
	year={2011},
	publisher={Elsevier}
}

@book{risken1996fokker,
	title={Fokker-planck equation},
	author={Risken, Hannes and Risken, Hannes},
	year={1996},
	publisher={Springer}
}

@book{tabar2019analysis,
	title={Analysis and data-based reconstruction of complex nonlinear dynamical systems},
	author={Tabar, Rahimi},
	volume={730},
	year={2019},
	publisher={Springer}
}

@article{gorjao2019kramersmoyal,
	title={kramersmoyal: Kramers--Moyal coefficients for stochastic processes},
	author={Gorj{\~a}o, Leonardo Rydin and Meirinhos, Francisco},
	journal={arXiv preprint arXiv:1912.09737},
	year={2019}
}

@article{rinn2016langevin,
	title={The Langevin approach: An R package for modeling Markov processes},
	author={Rinn, Philip and Lind, Pedro G and W{\"a}chter, Matthias and Peinke, Joachim},
	journal={arXiv preprint arXiv:1603.02036},
	year={2016}
}

@article{lin2023discontinuous,
	title={Discontinuous Jump Behavior of the Energy Conversion in Wind Energy Systems},
	author={Lin, Pyei Phyo and W{\"a}chter, Matthias and Tabar, M Reza Rahimi and Peinke, Joachim},
	journal={P R X Energy},
	volume={2},
	pages={033009},
	year={2023},
	publisher={APS}
}

@inproceedings{lind2014reconstructing,
	title={Reconstructing the intermittent dynamics of the torque in wind turbines},
	author={Lind, Pedro G and W{\"a}chter, Matthias and Peinke, Joachim},
	booktitle={Journal of Physics: Conference Series},
	volume={524},
	number={1},
	pages={012179},
	year={2014},
	organization={IOP Publishing}
}

@article{gottschall2008improve,
	title={How to improve the estimation of power curves for wind turbines},
	author={Gottschall, Julia and Peinke, Joachim},
	journal={Environmental Research Letters},
	volume={3},
	number={1},
	pages={015005},
	year={2008},
	publisher={IOP Publishing}
}

@article{mucke2015langevin,
	title={Langevin power curve analysis for numerical wind energy converter models with new insights on high frequency power performance},
	author={M{\"u}cke, Tanja A and W{\"a}chter, Matthias and Milan, Patrick and Peinke, Joachim},
	journal={Wind Energy},
	volume={18},
	number={11},
	pages={1953--1971},
	year={2015},
	publisher={Wiley Online Library}
}

@article{nadaraya1964estimating,
	title={On estimating regression},
	author={Nadaraya, Elizbar A},
	journal={Theory of Probability \& Its Applications},
	volume={9},
	number={1},
	pages={141--142},
	year={1964},
	publisher={SIAM}
}

@article{watson1964smooth,
	title={Smooth regression analysis},
	author={Watson, Geoffrey S},
	journal={Sankhy{\=a}: The Indian Journal of Statistics, Series A},
	pages={359--372},
	year={1964},
	publisher={JSTOR}
}

@article{lamouroux2009kernel,
	title={Kernel-based regression of drift and diffusion coefficients of stochastic processes},
	author={Lamouroux, David and Lehnertz, Klaus},
	journal={Physics Letters A},
	volume={373},
	number={39},
	pages={3507--3512},
	year={2009},
	publisher={Elsevier}
}

@article{kleinhans2007maximum,
	title={Maximum likelihood estimation of drift and diffusion functions},
	author={Kleinhans, David and Friedrich, Rudolf},
	journal={Physics Letters A},
	volume={368},
	number={3-4},
	pages={194--198},
	year={2007},
	publisher={Elsevier}
}

@article{kleinhans2012estimation,
	title={Estimation of drift and diffusion functions from time series data: A maximum likelihood framework},
	author={Kleinhans, David},
	journal={Physical Review E},
	volume={85},
	number={2},
	pages={026705},
	year={2012},
	publisher={APS}
}

@article{garcia2017nonparametric,
	title={Nonparametric estimation of stochastic differential equations with sparse Gaussian processes},
	author={Garcia, Constantino A and Otero, Abraham and Felix, Paulo and Presedo, Jesus and Marquez, David G},
	journal={Physical Review E},
	volume={96},
	number={2},
	pages={022104},
	year={2017},
	publisher={APS}
}

@article{milan2013turbulent,
	title={Turbulent character of wind energy},
	author={Milan, Patrick and W{\"a}chter, Matthias and Peinke, Joachim},
	journal={Physical review letters},
	volume={110},
	number={13},
	pages={138701},
	year={2013},
	publisher={APS}
}
\end{document}